\documentclass[aps,prb,twocolumn,superscriptaddress,preprintnumbers]{revtex4-2}

\usepackage{graphicx,epstopdf} 
\usepackage{amsmath}
\usepackage{tabu}
\usepackage{caption}
\usepackage{subcaption}
\usepackage{ragged2e}
\usepackage{xcolor}
\usepackage{multirow}
\usepackage{soul}
\usepackage[normalem]{ulem}
\usepackage[colorlinks=true, linkcolor=blue, citecolor=blue, urlcolor=blue]{hyperref}

\newcommand{\LCGO}{CuLa$_2$Ge$_2$O$_8$}

\newcommand{\be}{\begin{equation}}
\newcommand{\ee}{\end{equation}}
\newcommand{\bea}{\begin{eqnarray}}
\newcommand{\eea}{\end{eqnarray}}
\newcommand{\bse}{\begin{subequations}}
\newcommand{\ese}{\end{subequations}}

\begin{document}

\title{Crystal growth and magnetic properties of spin-$1/2$ distorted triangular lattice antiferromagnet CuLa$_2$Ge$_2$O$_8$}

\author{S. Thamban}
\email{}
\affiliation{Institut für Festkörperphysik, Technische Universität Berlin, Hardenbergstrasse 36, 10623 Berlin, Germany}
\affiliation{Helmholtz-Zentrum Berlin für Materialien und Energie, Hahn-Meitner-Platz 1, 14109 Berlin, Germany}

\author{C. Aguilar-Maldonado}
\affiliation{Helmholtz-Zentrum Berlin für Materialien und Energie, Hahn-Meitner-Platz 1, 14109 Berlin, Germany}

\author{S. Chillal}
\affiliation{Helmholtz-Zentrum Berlin für Materialien und Energie, Hahn-Meitner-Platz 1, 14109 Berlin, Germany}

\author{R. Feyerherm}
\affiliation{Helmholtz-Zentrum Berlin für Materialien und Energie, Hahn-Meitner-Platz 1, 14109 Berlin, Germany}

\author{K. Prokeš}
\affiliation{Helmholtz-Zentrum Berlin für Materialien und Energie, Hahn-Meitner-Platz 1, 14109 Berlin, Germany}

\author{A. J. Studer}
\affiliation{Australian Nuclear Science and Technology Organisation (ANSTO), Lucas Heights, NSW 2234, Australia}

\author{D. Abou-Ras}
\affiliation{Helmholtz-Zentrum Berlin für Materialien und Energie, Hahn-Meitner-Platz 1, 14109 Berlin, Germany}

\author{K. Karmakar}
\affiliation{Helmholtz-Zentrum Berlin für Materialien und Energie, Hahn-Meitner-Platz 1, 14109 Berlin, Germany}

\author{A. T. M. N. Islam}
\affiliation{Helmholtz-Zentrum Berlin für Materialien und Energie, Hahn-Meitner-Platz 1, 14109 Berlin, Germany}

\author{ B. Lake}
\email{}
\affiliation{Institut für Festkörperphysik, Technische Universität Berlin, Hardenbergstrasse 36, 10623 Berlin, Germany}
\affiliation{Helmholtz-Zentrum Berlin für Materialien und Energie, Hahn-Meitner-Platz 1, 14109 Berlin, Germany}

\begin{abstract}	
CuLa$_2$Ge$_2$O$_8$ forms a distorted triangular lattice of quantum spin-1/2 Cu$^{2+}$ ions. A crystal growth method was developed using the traveling-solvent floating zone technique resulting in the synthesis of a large single crystal (4 mm$\times$4 mm$\times$10 mm). The crystal was characterized with regard to phase purity and crystallinity using powder X-ray diffraction, energy dispersive X-ray analysis and Laue diffraction, and found to be of excellent quality. The magnetic properties were characterized using dc-susceptibility, magnetization, and heat capacity measurements which revealed weak magnetic frustration with long-range magnetic order occurring below $T_N=1.14(1)$~K. The magnetic structure determined using neutron powder diffraction is a commensurate, noncollinear antiferromagnetic, different from the 120$^{\circ}$ order of an equilateral triangular antiferromagnet. The ordered moments lie in the {\bf bc}-plane, with components $m_b=0.50(3)$~$\mu_{B}$ and $m_c= 0.73(5)$~$\mu_{B}$ along the {\bf b}- and {\bf c}-axes respectively, giving a total ordered moment of $M_{total}$= 0.89(6)$\mu_{B}/$Cu$^{2+}$ at 20~mK.
\end{abstract}

\maketitle
\section{INTRODUCTION}
Over the last few decades, geometrically frustrated quantum magnets have been attracting widespread research interest among condensed matter physicists because of their possible realization of unconventional magnetism manifested as exotic ground states and excitations \cite{balents2010,sachdev2008quantum,harris1997geometrical,khatua2022signature,balz2016physical}, for example a possible quantum spin liquid (QSL) state was found in Na$_2$BaCo(PO$_4$)$_2$\cite{Zhong14505}, quantum spin-ice in Pr$_2$Hf$_2$O$_7$\cite{sibille2018}, a magnetization plateau in Ba$_3$CoSb$_2$O$_9$\cite{shirata2012experimental} and exotic excitations in YbMgGaO$_4$\cite{paddison2017continuous}.

Geometrical frustration arises due to spin arrangements in specific types of lattices such as triangular\cite{collins1997review}, pyrochlore\cite{gardner2010magnetic}, kagom\'{e}\cite{syozi1951statistics}, etc. The simplest among these is the two-dimensional (2D) triangular lattice with antiferromagnetic (AFM) nearest neighbor interactions; however, many of its physical properties are still not fully understood. Magnetic ordering is typically of helical form, with adjacent spins separated by an angle of 120$^{\circ}$. However, when antiferromagnetic next-nearest-neighbor interactions are introduced, the excitations within the system are perturbed. This interaction shifts magnetic ordering from a helical configuration to a collinear structure through a quantum critical region\cite{wheeler2009spin,chubukov1992order}. To verify theoretical predictions and to search for new ground states in frustrated quantum magnets, it is essential to search for new types of geometrically frustrated compounds. Here, we investigate the 2D AFM triangular lattice compound CuLa$_2$Ge$_2$O$_8$ first reported in Ref.~\cite{cho2017properties}, where Cu$^{2+}$ is the magnetic ion with quantum spin-1/2. 

\begin{figure}[h]
 \includegraphics[scale=0.9]{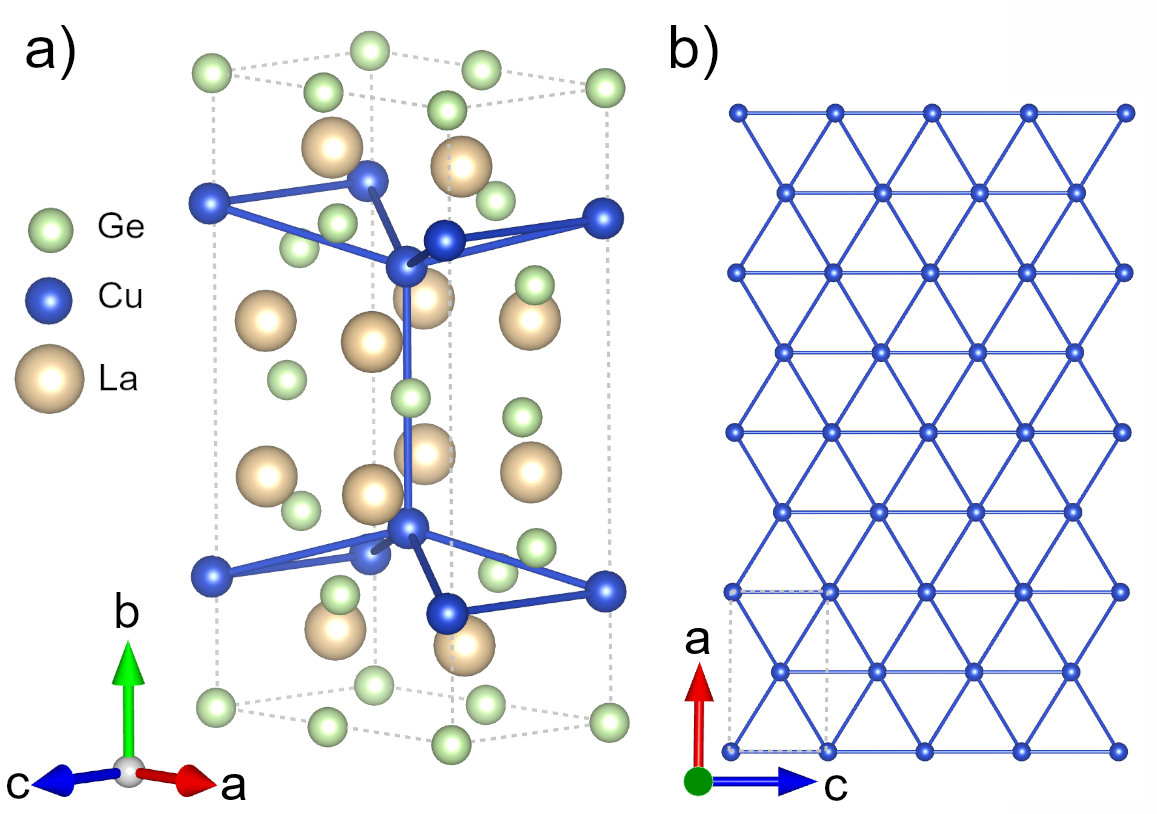}
 \captionsetup{justification=justified, singlelinecheck=false}
\caption{\label{fig:Fig1}
    \justifying
    The crystal structure of CuLa$_2$Ge$_2$O$_8$ where the Oxygen atoms are removed for clarity. The Cu$^{2+}$ ions are displayed as blue spheres, Germanium atoms are represented as green spheres and Lanthanum as yellow spheres, b) The Cu$^{2+}$ ions form distorted triangular layers within the {\bf ac}-plane where the distance between the neighbouring Cu$^{2+}$ ions varies from 5.14 to 5.17~\AA. The smallest interlayer distance is 6.37~\AA.
}
\end{figure}

The structural and magnetic properties of CuLa$_2$Ge$_2$O$_8$ such as dc-susceptibility, magnetization, and heat-capacity  were reported by Cho {\em et.\ al.}\cite{cho2017properties}. The results indicate that it crystallizes in the monoclinic crystal system with the $I1m1$ space group. The crystal structure of CuLa$_2$Ge$_2$O$_8$ contains distorted CuO$_4$ plaquettes, GeO$_4$ tetrahedra and GeO$_5$ trigonal bipyramids. The distorted CuO$_4$ plaquettes can interact with each other to form a distorted triangular lattice in the {\bf ac}-plane as shown in Fig.~\ref{fig:Fig1}. The previous bulk characterization revealed that the system shows a broad maximum at $T_{\mathrm{max}} \approx1.40(4)$~K and a Curie-Weiss temperature $\theta_{\mathrm{CW}}$ of -5.7(4)~K, while long-range magnetic order occurs at $T_{\mathrm{LRO}} = 0.93(2)$~K giving the ratio $\theta_{\mathrm{CW}} / T_{\mathrm{LRO}}=6.1(4)$ and implying that it is frustrated. Heat capacity measurments gave a similar N\'{e}el temperature of $T_\mathrm{N}=1.09(4)$~K \cite{cho2017properties}. Cho {\em et.~al.} performed all their measurements on single crystals  grown by the flux method with dimensions less than a millimeter limiting the precision of their results. This has motivated us to grow larger crystals for further detailed investigation of frustrated quantum magnetism using advanced measuring techniques, including neutron diffraction.

Here, we have grown the first large millimeter-sized single crystals of CuLa$_2$Ge$_2$O$_8$ by using the traveling solvent floating zone (TSFZ) method. These allowed us to perform a thorough investigation of its structural and magnetic properties including a determination of its magnetic structure.

\section{\textbf{EXPERIMENTAL DETAILS}}

\subsection{\textbf{Synthesis}}

Both powder and single crystals of CuLa$_2$Ge$_2$O$_8$ were synthesized and investigated experimentally at the Core Lab Quantum Materials at Helmholtz Zentrum Berlin f\"{u}r Materialien und Energie (HZB) Germany. The powder samples were made by solid state reaction.
The single crystal was grown by the TSFZ technique\cite{triboulet2015crystal, wolff1974crystal,hurle1994rw,koohpayeh2016single}. An optical furnace (CSI FZ-T10000-H-VI-VP, Crystal Systems, Inc., Japan) equipped with four ellipsoidal mirrors and four 500W tungsten halide lamps were chosen for the growth process.

For the powder synthesis high-purity powders of CuO (99.9985$\%$, Alfa Aesar),  La$_2$O$_3$ (99.995$\%$, Alfa Aesar) and GeO$_2$ (99.9985$\%$, Alfa Aesar) were mixed in the 1:1:2 molar ratio. An initial thermal treatment was performed on La$_2$O$_3$ at $T=700$~$^{\circ}$C for 12 h to remove H$_2$O. The stoichiometric powder was first sintered for 12 h at 850~$^\circ$C, and a subsequent sintering was performed at 950~$^\circ$C and finally at 1000~$^\circ$C for 24 h each. In between each sintering, the powder was ground thoroughly. The resulting polycrystalline powder was used for the X-ray and neutron powder diffraction measurements discussed later. 

For the crystal growth, the feed rod was obtained from the pure powder which was packed into a cylindrical rubber tube and pressed hydrostatically up to 2000~bar in a cold isostatic press. This feed rod was then sintered at 1100~$^\circ$C for 24 h. A dense cylindrical rod with $\sim 5$~mm diameter and $\sim 60$~mm length was confirmed to be a pure CuLa$_2$Ge$_2$O$_8$ phase by means of powder X-ray diffraction. 

\begin{figure}[h]
\includegraphics[scale=0.32]{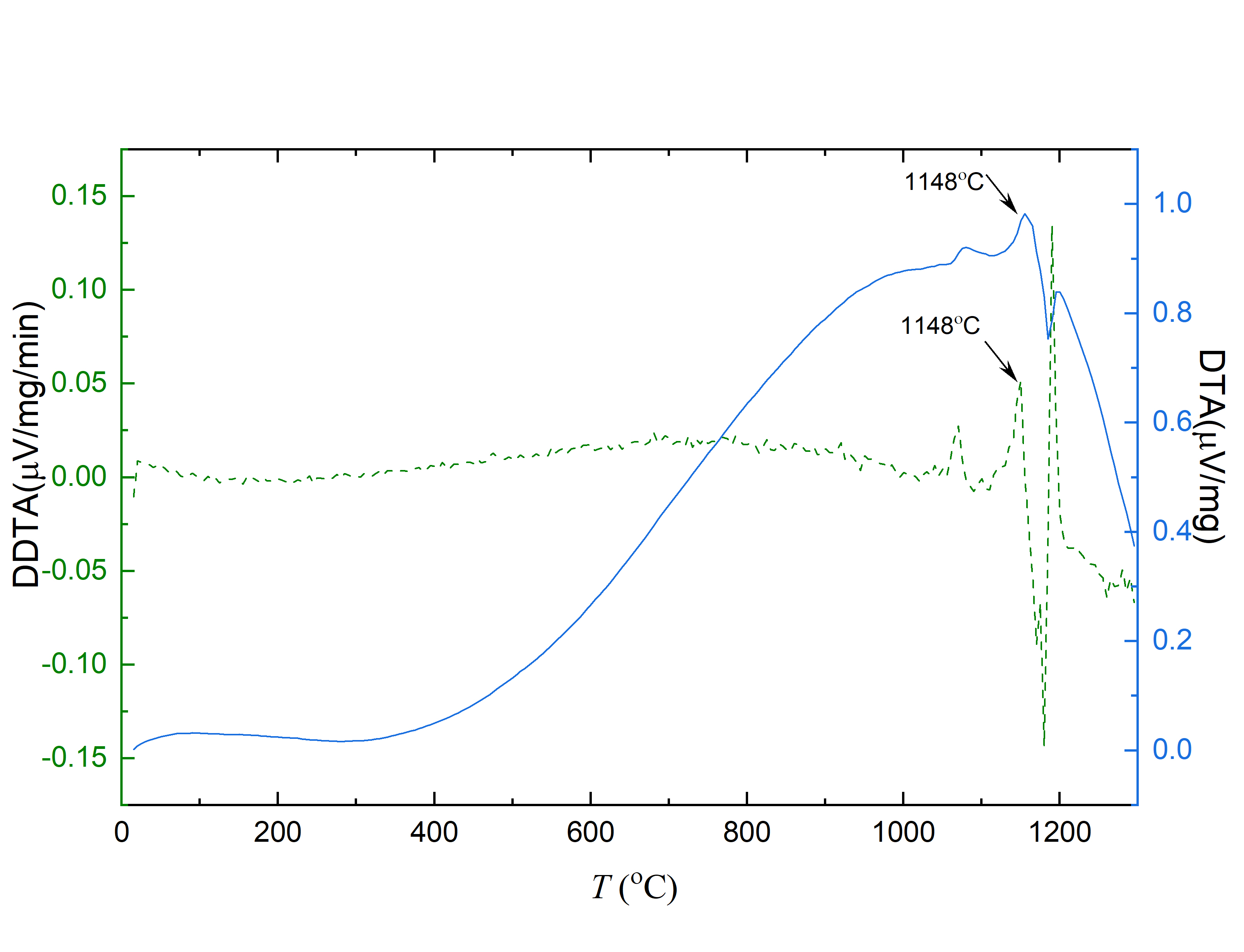}
 \captionsetup{justification=justified, singlelinecheck=false}
 \caption{\label{fig:tga}\justifying Differential Thermal Analysis (DTA) (blue line, right axis) measured on phase pure poly-crystalline CuLa$_2$Ge$_2$O$_8$. The compound decomposes into La$_2$Ge$_2$O$_7$ and Cu$_2$O at a temperature of 1148~$^\circ$C before melting at a temperature range 1170-1180~$^\circ$C. The DDTA shown on the left axis is the derivative of DTA, where black arrows indicates the decomposition temperature.}
\end{figure}

\begin{figure}[h]
 \includegraphics[scale=0.38]{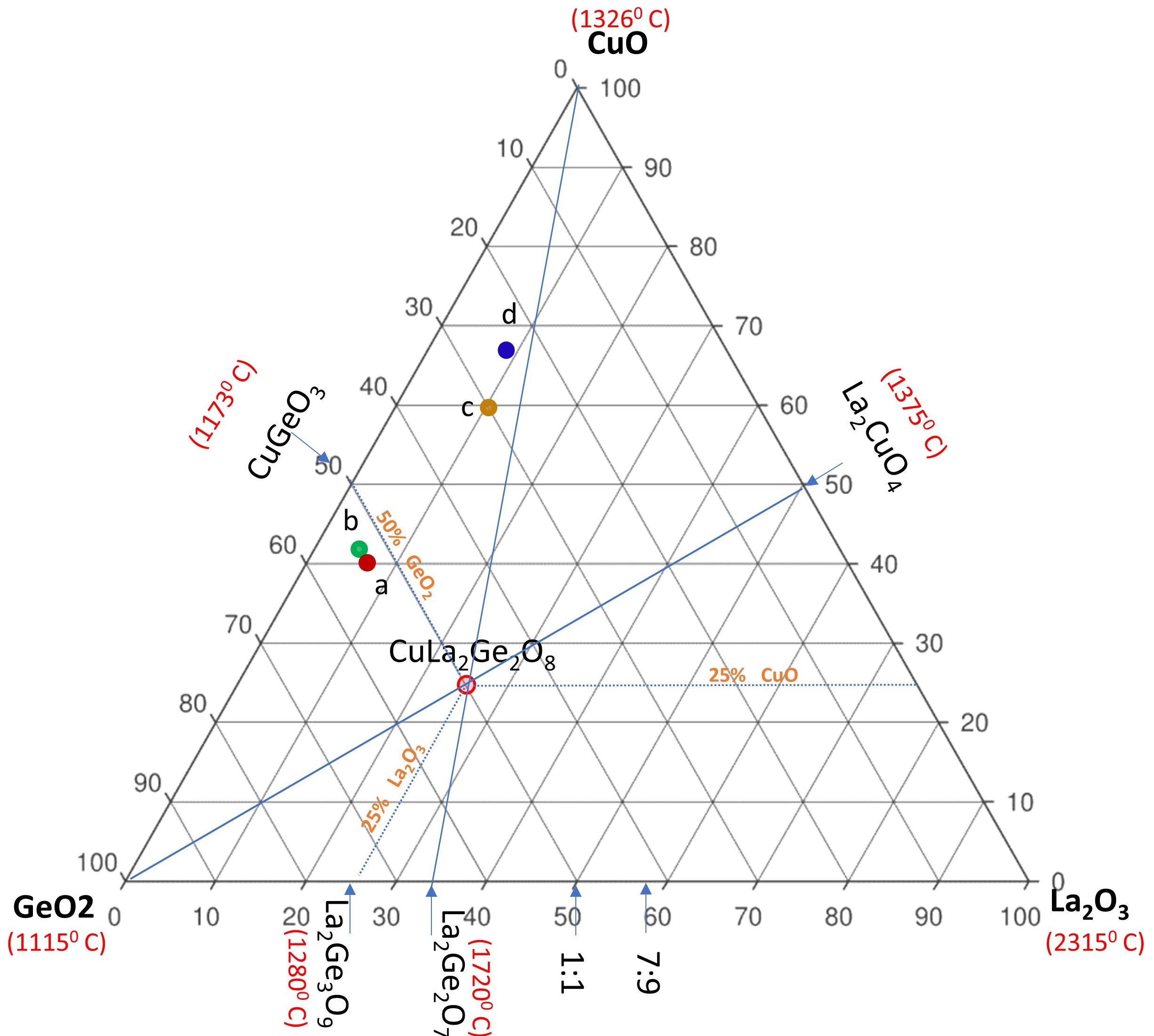}\\
 \captionsetup{justification=justified, singlelinecheck=false}
 \caption{\label{fig:PhaseDiagram}\justifying Phase diagram for the GeO$_2$-La$_2$O$_3$-CuO system. The colored dots (a,b,c,d) are the different solvent compositions tried during the crystal growth experiments.}
\end{figure}

A differential thermal analysis (DTA) of the powder showed that CuLa$_2$Ge$_2$O$_8$ decomposes in to La$_2$Ge$_2$O$_7$ and Cu$_2$O at 1148~$^\circ$C before melting (Fig.~\ref{fig:tga}). Therefore a suitable self-flux is desired for a successful growth to reduce the melting temperature. Since no report was found in the literature of the temperature-composition phase diagram for CuLa$_2$Ge$_2$O$_8$, an investigation of the two-component phase diagrams of CuO-La$_2$O$_3$\cite{Qiao,oka1987phase} CuO-GeO$_2$ \cite{CuO}, and GeO$_2$-La$_2$O$_3$ \cite{Bondar} was carried out to get insight into the possible phase relations in CuLa$_2$Ge$_2$O$_8$. After a careful analysis, four different solvent compositions of CuO, La$_2$O$_3$, and GeO$_2$ in the molar ratios (a) 6:1:8, (b) 8:1:10, (c) 6:1:3 and (d) 8:1:3 were tried. The first two have an excess in CuGeO$_3$ and latter an excess of CuO. The choice of the solvents are shown in ternary phase diagram of CuO-GeO$_2$-La$_2$O$_3$ (Fig.~\Ref{fig:PhaseDiagram}). 

Each of the solvent compositions was prepared by solid state reaction at 800~$^\circ$C for 12 h. For TSFZ growth a solvent disk of about 0.4-0.5~g was first attached to one end of the feed rod by melting the end of the rod in the optical floating zone furnace. Crystal growth was initiated by remelting the solvent part and seeding the high-temperature melt onto the seed rod. Initially, a portion of the feed rod was used as the seed rod then subsequently, a crystal from a previous growth was used to avoid random nucleation. To have sufficient mixing within the melt the two rods were counter-rotated at a certain speed. Since the growth was planned at temperatures where decomposition of CuO to Cu$_2$O and evaporation of copper oxides was expected, the optimization of the growth atmosphere was crucial for stable growth. For this reason, different growth atmospheres of high-pressure oxygen between 0.2 to 0.5~MPa and also a mixture of O$_2$ and Ar in high pressure (0.3~MPa) were tested. Different growth speeds were also tried from 0.3 to 1.0~mm/h.

After the growth, different parts of the rod were carefully checked using a polarized optical microscope and Laue diffraction to find single crystal parts along the length. Once a single crystal part was identified, it was then checked for phase purity by Powder X-ray diffraction at room temperature (Bruker D8) using Cu K$_{\alpha1}$ radiation ($\lambda$=1.5406 \AA)  and Single Crystal X-ray diffraction at room temperature (Bruker Kappa-APEX2) to obtain the exact structure of the crystal ($\lambda$=0.70932 \AA). A piece of the single crystal was polished and checked by means of energy-dispersive X-ray analysis (EDX) spectroscopy in a scanning electron microscope (SEM) for any impurity phases or grain boundaries.

\subsection{\textbf{Magnetic, thermodynamic and neutron measurements}}

The magnetic and physical bulk properties were explored at the Core Lab Quantum Materials, HZB. The susceptibility and magnetization along all the crystallographic axes were measured using a MPMS3 SQUID magnetometer (Quantum Design). Magnetic fields up to 7~T were applied, and temperatures down to 0.4~K were achieved using a $^3$He insert. Heat capacity measurements were performed using a Quantum Design Physical property measurement system (PPMS) in the temperature range of 0.4~K to 200~K and magnetic fields up to 9~T. 

The neutron diffraction pattern of \LCGO\ was collected using the high-intensity neutron diffractometer Wombat \cite{studer2006wombat}, located in the OPAL Neutron Guide Hall at the Australian Nuclear Science and Technology Organisation, (ANSTO) near Sydney, Australia. The diffraction measurement was performed on a powder sample of mass 10.0(1)~g placed inside a copper can, which was cooled in a dilution refrigerator. The diffraction patterns were collected for 12 h between the temperature $T = 20$~mK and 3~K, using an incident wavelength of 4.08(1)~\AA~ and a detector coverage of 2$\theta = 109^{\circ}$ with a step size of 0.125$^{\circ}$. From these data sets the low-T nuclear and magnetic structures were refined.

\section{\textbf{RESULT AND DISCUSSION}}

\subsection{\textbf{Crystal growth}}

\begin{figure}[h]
 \includegraphics[scale=0.4]{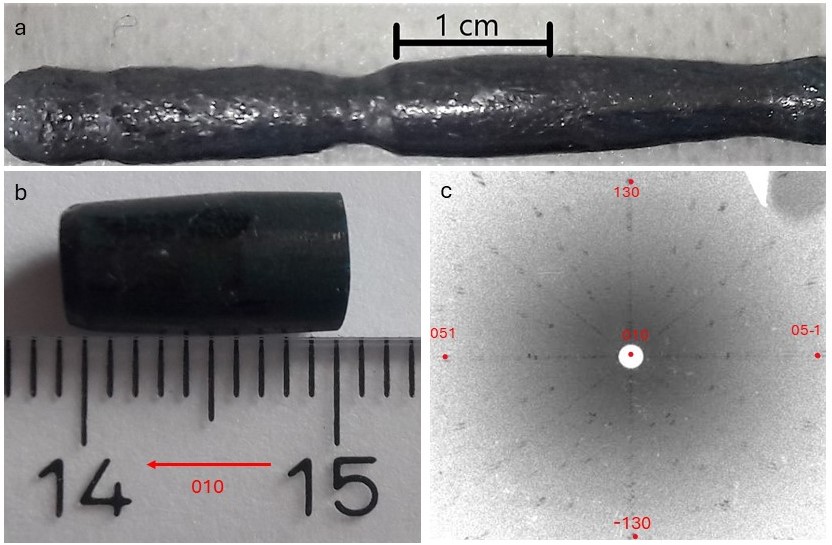}\\
 \captionsetup{justification=justified, singlelinecheck=false}
 \caption{ \label{fig:crystal}\justifying (a) Photograph of  the as-grown crystal from the optimized TSFZ method, (b) Single crystalline part from the (a), (c) X-ray Laue diffraction image of the single crystal part shown in (b) with [010]-axis parallel to the incident X-ray beam. The growth direction is slightly off from the [010]-axis (by $\sim5^\circ$).}
\end{figure}

\begin{table}[htbp]
\setlength{\tabcolsep}{0.2cm}
\caption{\label{tab:growth}Optimized growth parameters}
\begin{tabular}{l|l}
\hline \hline		
Feed/seed rods       & CuLa$_2$Ge$_2$O$_8$     \\[0.1cm]
Atmosphere & Ar:O$_2$; 1:1 at 0.3~MPa \\[0.1cm]
Solvent   & CuO:La$_2$O$_3$:GeO$_2$: 6:1:8 \\[0.1cm]
Growth rate  &0.35~mm/hr \\[0.1cm]
Shafts rotation speed  & 25~rpm\\[0.1cm]
Lamp power  &500~W \\[0.1cm]	
\hline
\hline
\end{tabular}
\end{table}

\begin{figure}[htbp]
 \includegraphics [width=\linewidth] {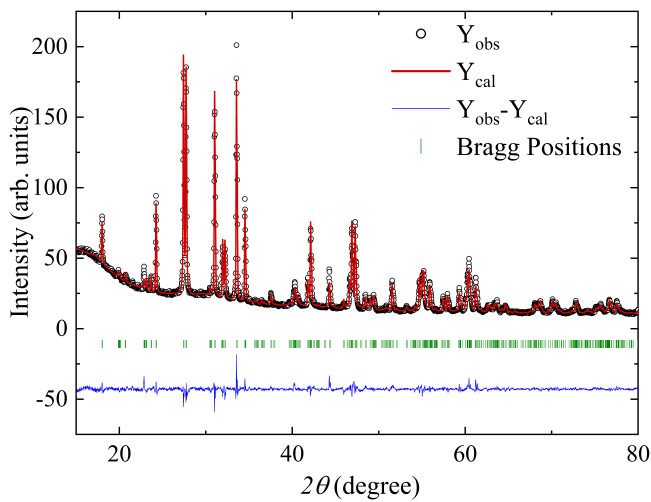}\\
 \captionsetup{justification=justified, singlelinecheck=false}
 \caption{\label{fig:Fig4}\justifying Powder XRD pattern (open circles) measured on the crushed single crystal at room temperature. The red solid line represents the Rietveld refinement, with the green vertical bars showing the fitted Bragg peak positions, and the lower blue solid line represents the difference between the observed and calculated intensities.}
\end{figure}

\begin{figure*}
 \includegraphics [scale=.55] {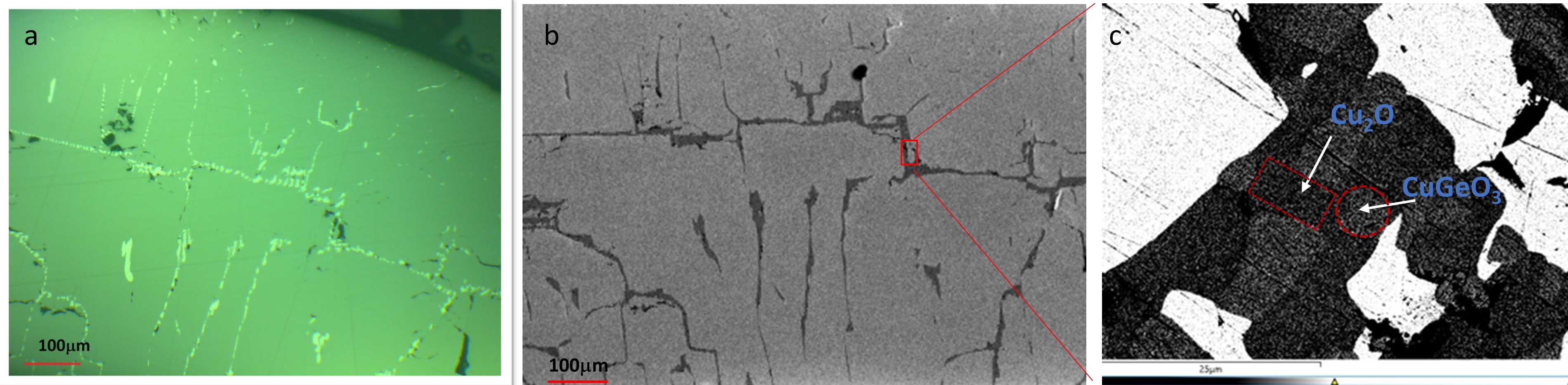}\\
 \captionsetup{justification=justified, singlelinecheck=false}
 \caption{\label{fig:microscope}\justifying (a) Polarized optical microscope image of a small part of the crystal, indicating inclusions or impurities. (b) The same crystal portion imaged by a scanning electron microscope. An enlarged image of the region indicated by the red box shown in (c), which clearly distinguishes the impurities. EDX analysis indicates the impurities are Cu$_2$O and CuGeO$_3$ as labelled in (c).}
\end{figure*}

Several growth processes were attempted by the TSFZ method using different solvent compositions labeled (a) to (d) in Fig.~\ref{fig:PhaseDiagram} as well as employing different growth atmospheres with a growth speed of 1.0~mm/h. Solvents (c) and (d) were found to require a higher lamp power to melt the rod compared to solvents (a) and (b). The melt zone for (c) and (d) was soon lost, and these growths were stopped after 1-2 h. The presence La$_2$Ge$_2$O$_7$ in the initial part of growth was confirmed by powder X-ray diffraction, indicating that the temperature was too high. On the other hand, for solvent (a) the required lamp power was the same, while for solvent (b) the required power was lower than that required for (c) and (d). The growth of (a) and (b) seemed stable for the first 5-7 h, and then the melt zone started to become thinner, and the lamp power had to be increased to maintain growth; this eventually led to the interruption of growth. The surface of the as-grown crystals was rough. Higher amounts of CuGeO$_3$ as an impurity were found in the initial part of growth when using solvent (b). Therefore, solvent (a) was considered to be closest to the optimum. 

An atmosphere of Ar mixed with O$_2$ at high pressure was found to increase thermal conductivity and help improve the quality of the crystal. The growth speed was also reduced to between 0.3 to 0.5~mm/h. A stable growth lasting several days could be achieved using the growth parameters summarized in Table~\Ref{tab:growth}. The as-grown single crystal is shown in Fig.~\Ref{fig:crystal}(a). Several different cross sections of the as-grown rod were analyzed using an optical microscope finding a 4~mm$\times$4~mm$\times$10~mm in size single crystal as shown in Fig.~\Ref{fig:crystal}(b). A Laue diffraction pattern along the growth direction of this crystal part is shown in Fig.~\Ref{fig:crystal}(c) .

\subsection{\textbf{Characterization}}

 To analyse the crystal structure and composition, a small part of it was crushed into a powder and measured using X-ray powder diffraction (XRD) at room temperature. Several compounds from the same family can be found in the literature.  CuNd$_2$Ge$_2$O$_8$\cite{campa1995cund2ge2o8} is reported to have the $C1m1$ space group with cell parameters $a = 9.846(2)$~\AA, $b = 15.335(5)$~\AA, $c = 8.336(1)$~\AA~ and $\beta = 148.48(2)\,^{\circ}$. Meanwhile CuRE$_2$Ge$_2$O$_8$ (RE = Y, La)\cite{cho2017properties} and (RE=Pr, Nd, Sm, Eu)\cite{cho2017CuR} have been reported to have the space group $I1m1$, with cell parameters $a = 8.587(2)$~\AA, $b = 15.565(6)$~\AA, $c = 5.199(1)$~\AA~ and $\beta = 89.30(1)$ deg. A detailed structural analysis of our diffraction pattern was performed, and both space groups were taken into account for the Rietveld refinement. Figure~\Ref{fig:Fig4} shows the Rietveld refinement performed with FullProf software\cite{Carvajal55}. The red line indicates the calculated pattern, and the green bars indicate the Bragg positions. The best fit confirms a monoclinic structure (space group $I1m1$, \emph{Z}=4)  with lattice parameters $a = 8.574(5)$~\AA, $b = 15.562(2)$~\AA, $c = 5.192(4)$~\AA , $\beta = 89.348(1)$ deg. The refinement yielded a  goodness-of-fit  $\chi^2 \simeq 2.95 $ in agreement with previous reports\cite{cho2017properties}. There were no impurities detected in the powder X-ray diffraction pattern of the crystal. The CuGeO$_3$ and Cu$_2$O found in the EDX measurement discussed later were not detectable.

As a further check of the quality and structure of the single crystal of CuLa$_2$Ge$_2$O$_8$, a single crystal X-ray diffraction experiment was carried out on a small piece at room temperature that confirmed the space group to be $I1m1$. The refinement was performed using \texttt{FullProf} and the atomic coordinates are given in Table \ref{tab:Atomic}.
\begin{table}[htbp]
\setlength{\tabcolsep}{0.2cm}
\captionsetup{justification=justified, singlelinecheck=false}
\caption{\label{tab:Atomic}\justifying Rietveld refinement of the atomic coordinates from the single crystal X-ray diffraction data of CuLa$_2$Ge$_2$O$_8$ at room temperature. The lattice cell parameters $a = 8.582(1)$~\AA, $b = 15.549(1)$~\AA, $c = 5.185(2)$~\AA , $\beta = 89.3(1)$ were similar to the powder refinement with a goodness-of-fit of 1.67.}	
\begin{tabular}{ccccc}
\hline \hline
Atoms       &$x$     &$y$    &$z$      &$U_{\rm eq}$ \\
\hline
La1& 0.7312(2)& 0.12341(3)& 0.5220(4)& 0.00818(8)\\[0.1cm]
La2& 0.2193(2)& 0.12041(4)& 0.5797(4)& 0.00949(9)\\[0.1cm]
Ge1& 0.0076(2)& 0.5& 0.5546(4)& 0.00496(15)\\[0.1cm]
Ge2& 0.5067   & 0.5& 0.5246& 0.0087(2)\\[0.1cm]
Ge3& 0.0252(3)& 0.29307(8)& 0.5003(4)& 0.00872(14)\\[0.1cm]
Cu1& 0.4970(3)& 0.29365(10)& 0.5195(4)& 0.00895(16)\\[0.1cm]
O1 & 0.8474(12)& 0.5& 0.840(2)& 0.0058(10)\\[0.1cm]
O2 & 0.1712(7)& 0& 0.8111(12)&-0.0031(4)\\[0.1cm]
O3 & 0.8535(7)& 0& 0.8515(13)&-0.0013(5)\\[0.1cm]
O4 & 0.1790(8)& 0.5& 0.8490(14)&-0.0002(5)\\[0.1cm]
O5 & 0.5121(8)& 0.0914(4)& 0.8771(14)& 0.0048(6)\\[0.1cm]
O6 & 0.9945(8)& 0.3259(4)& 0.8634(13)& 0.0045(6)\\[0.1cm]
O7 & 0.4999(13)& 0.3998(6)& 0.693(2)& 0.0151(13)\\[0.1cm]
O8 & 0.9891(8)& 0.1716(4)& 0.7410(13)& 0.0053(6)\\[0.1cm]
O9 & 0.3662(10)& 0.2396(5)& 0.8070(16)& 0.0093(10)\\[0.1cm]
O10& 0.7063(14)& 0.2281(8)& 0.848(3)& 0.0193(18)  \\[0.1cm]      
\hline
\hline
\end{tabular}
\end{table}

\begin{figure}[htbp]
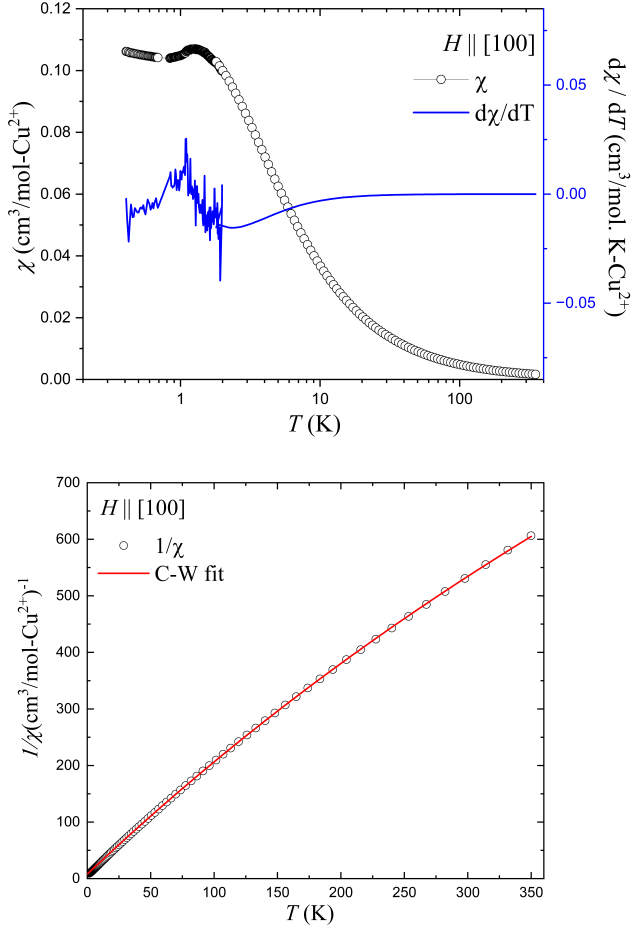

    \centering
    \includegraphics[width=0.5\textwidth]{susceptibility_new.jpg}
    \includegraphics[width=0.46\textwidth]{invsusceptibility_new.jpg}
    \captionsetup{justification=justified, singlelinecheck=false}
    \caption{\label{fig:chi}\justifying Upper panel shows the $\chi(T)$ (left hand axis) and d$\chi$/d$T$ (right hand axis) measured in a field of $\mu_0 H= 1$~T along the [100] direction over the temperature range 0.35~K to 340~K. A clear 3D long range magnetic order at 1.1 K is observed from the d$\chi$/d$T$ curve. Lower panel shows the $1/(\chi)$ vs $T$  measured at $H= 1$~T for crystallographic (100) direction from 0.35~K to 350~K. The solid line represents the CW fit using Eq.~\eqref{cw}.}
\end{figure}

A thin section was cut out from the grown crystal using a diamond cutter and dry polished using polishing paper and lubricant liquid (of various micrometer sizes in order to get a smooth surface). This polished part was used to identify the presence of impurity phases or inclusions or grain boundaries under a polarized optical microscope as shown in Fig.~\ref{fig:microscope}(a)). In Fig.~\ref{fig:microscope}(a) we can identify different phase contrasts which were further examined using a SEM equipped with EDX (see Fig.~\ref{fig:microscope}(b,c)).  EDX analysis revealed that the stoichiometric ratio expected for the chemical formula CuLa$_2$Ge$_2$O$_8$ over most of the cross section. At some points, there was a difference from the expected stoichiometric ratio (see Fig.~\ref{fig:microscope}(c)), suggesting a small volume fraction of secondary impurity phase present in the crystal. Using this average ratio, it was found that the possible impurity phases were CuGeO$_3$ and Cu$_2$O. From the above analyses, we can say that the synthesized crystal is a single crystal of CuLa$_2$Ge$_2$O$_8$, and  that the volume fraction of the impurity phases are altogether lower than 3\%. A small part of the crystal was used for thermodynamic and magnetic measurement along different crystallographic directions, which were aligned using an X-ray Laue diffractometer.

\subsection{Magnetization measurements} 

The direct current (DC) magnetic susceptibility $\chi(T)$ = $\frac{M(T)}{H}$ where $M(T)$ is the temperature-dependent magnetization, was measured in an applied field of $\mu_0H=1$~T along the three different crystallographic directions ([100], [010], and [001]) of the single crystal. The $\chi(T)$ versus $T$ along these different directions yielded similar results, only $H||$[100] is shown here (Fig.~\ref{fig:chi}). To extract the magnetic parameters, the $1/\chi(T)$ data was fitted in the temperature range 30~K to 240~K using Eq.~\eqref{cw}. 
\begin{equation}\label{cw}
\chi(T) = \chi_0 + \frac{C}{T - \theta_{\rm CW}},
\end{equation}
Where $\chi_0$ is the temperature-independent contribution consisting of the diamagnetic susceptibility ($\chi_{\rm core}$) of the core electron shells and Van-Vleck paramagnetic susceptibility ($\chi_{\rm VV}$) of the open shells of the Cu$^{2+}$ ions present in the sample. The second term in Eq.~\eqref{cw} is the Curie-Weiss (CW) law with the CW temperature ($\theta_{\rm CW}$) and Curie constant $C = N_{\rm A} \mu_{\rm eff}^2/3k_{\rm B}$, where $N_{\rm A}$ is the Avogadro number, $k_{\rm B}$ is the Boltzmann constant, $\mu_{\rm eff} = g\sqrt{S(S+1)}$$\mu_{\rm B}$ is the effective magnetic moment, $g$ is the Land$\acute{\rm e}$ $g$-factor, $\mu_{\rm B}$ is the Bohr magneton, and $S$ is the spin quantum number. The fit yields $\chi_0 \simeq 3.28 \times 10^{-4}$~cm$^3$/mol-Cu$^{2+}$, $C\simeq$ 0.468(1)~cm$^3$K/mol-Cu$^{2+}$ and $\theta_{\rm CW} \simeq -3.74(1)$~K. Using $C$, we have calculated the effective magnetic moment $\mu_{\rm eff}=1.93(1)\mu_B$. This value $\mu_{eff}$ corresponds to the Land$\acute{\rm e}$ $g$-factor of $g$= 2.23 which is slightly higher than the ideal $g$=2, expected for spin-$1/2$ ($\mu_{\rm eff}=1.732\mu_B$, assuming g=2). This deviation in  effective magnetic moment is consistent with the experimental observations for Cu$^{2+}$ compounds where the g-factors typically range from 2.1 to 2.3 \cite{Rnath2014,Arango2011,Guchhait_2021}. The negative $\theta_{\rm CW}$ value indicates that the dominant exchange couplings between Cu$^{2+}$ ions are antiferromagnetic. The derivative of $\chi(T)$ with respect to $T$ reveals a transition at $T_N=1.10(1)$~K showing that \LCGO\ orders antiferromagnetically (see upper panel of Fig~\ref{fig:chi}). The value of the N\'eel temperature is significantly lower than the Curie-Weiss temperature, giving a frustration index of $f=\theta_{\rm CW}/T_N = 3.38$ and suggesting the possible presence of weak frustration. 

\begin{figure}[htbp]
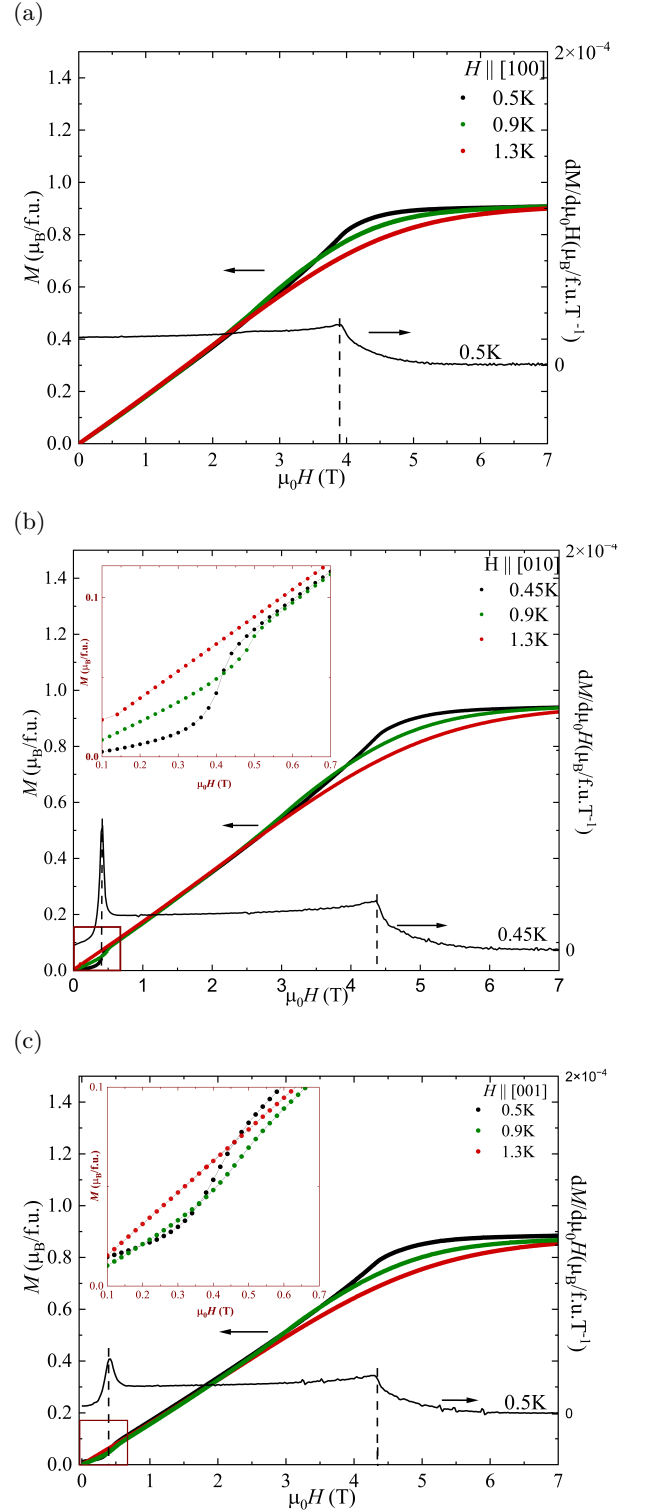

\begin{subfigure}{0.45\textwidth}
\captionsetup{justification=raggedright, singlelinecheck=false}
\caption{}
\includegraphics [scale=.29] {MH100.jpg}\\
\captionsetup{justification=justified, singlelinecheck=false}
\label{MH100}
\end{subfigure}\hfil 
\begin{subfigure}{0.45\textwidth}
\captionsetup{justification=raggedright, singlelinecheck=false}
\caption{}
\includegraphics [scale=.27] {MH010.jpg}\\
\label{MH010}
\end{subfigure}\hfil 
\begin{subfigure}{0.45\textwidth}
\captionsetup{justification=raggedright, singlelinecheck=false}
\caption{}
\includegraphics [scale=.29] {MH001.jpg}\\
\label{MH001}
\end{subfigure}\hfil 
\caption{\justifying  Magnetic isotherms ($M$ vs $H$) (left y-axis) and their field derivatives (right y-axis) were measured for (a) $H\parallel$[100], (b) $H\parallel$[010] and (c) $H\parallel$[001]. The measurements were performed at temperatures ranging from 0.4~K to 1.3~K with magnetic fields up to 7~T. For $H\parallel$[010] and $H\parallel$[001] at low fields and temperatures, the curves display a transition, as highlighted in the insets of (b) and (c). This transition is absent for $H\parallel$[100]. The vertical dashed black lines indicate the critical fields obtained from corresponding field derivatives at 0.5~K.}
\label{MH}
\end{figure}

To investigate the behavior of CuLa$_2$Ge$_2$O$_8$ in a magnetic field, the magnetization ($M$~versus~$H$) was measured at different temperatures for the three crystallographic directions (Fig.~\ref{MH}). In all cases, the magnetization increases approximately linearly with field, and saturation is achieved at a field of $\approx 3.8-4.5$~T. At lowest temperatures (0.45 - 0.5~K) an upward curvature can be observed when approaching the saturation field from below,  which is typical for low-dimensional magnets in the ordered state.\cite{parkinson2010introduction, bonner1964linear,blundell2001magnetism}. For the crystallographic directions [010] and [001] (Fig.~\ref{MH010} and \ref{MH001}) at the lowest temperature there is an anomaly as the magnetic field increases from zero with a sudden increase in the magnetization at $\approx 0.4$~T. This could indicate the presence of a spin-flop transition suggesting that the spins have components along the {\bf b}- and {\bf c}-axes. The sudden increase in magnetization is absent for [100] direction indicating that the ordered moments have no component along the {\bf a} axis below 0.4 T. 

The field derivative of M($H$) gives the field value of this transition as $\mu_0H_{SF}=0.42$~T. The field derivative can also be used to extract the saturation field where the spins all point parallel to the field direction. For $H\parallel$[100] at 0.5~K the fully polarized state occurs at $\mu_0H_{S}=3.87$~T, while for $H\parallel$[010] and $H\parallel$[001] the saturation fields are $\mu_0H_{S}=4.23$~T and $\mu_0H_{S}=4.33$~T, respectively.

\subsection{Heat capacity measurement} 
\begin{figure}
 \includegraphics [width=\linewidth] {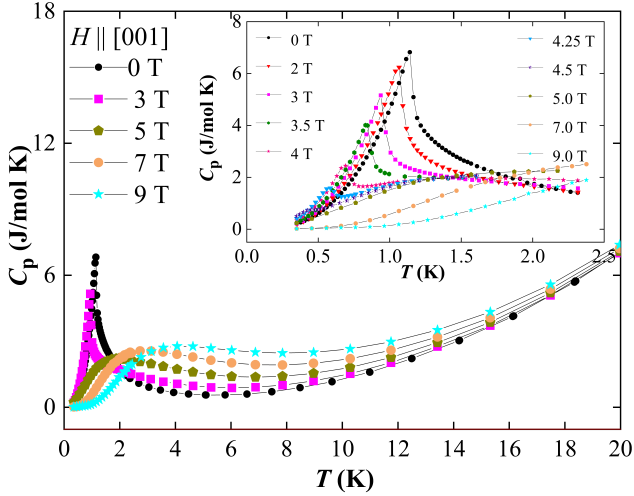}\\
 \captionsetup{justification=justified, singlelinecheck=false}
 \caption{\justifying Heat capacity data collected as a function of temperature from a single crystal of \LCGO ~for different fields applied along the crystallographic {\bf c}-axis. $\lambda$-anomaly , indicating the onset of antiferromagnetic order, shifts towards lower temperature on increasing field (see inset).}
 \label{Fig7(a)}
\end{figure}

Heat capacity $C_{\rm p}$ measurements were performed on single crystals of CuLa$_2$Ge$_2$O$_8$ as a function of temperature under different magnetic fields applied along the {\bf c}-axis in the temperature range from 0.4~K to 90~K as shown in Fig.~\ref{Fig7(a)}. The inset shows the low-temperature heat capacity from 0.3~K to 2.5~K. At $\mu_0 H = 0$~T, as temperature decreases the heat capacity also decreases down to $\approx 6$~K and then starts to increase. At $T_N=1.14(1)$~K, a sharp $\lambda$-anomaly occurs indicating the transition to three-dimensional long-range magnetic order. An additional broad feature on the high temperature side of the $\lambda$-anomaly suggest the presence of short-range magnetic order just above the transition. With increasing magnetic field, the transition moves to lower fields and is completely suppressed above 4.5~T in the saturated phase. Furthermore, the broad feature on the high temperature side of the $\lambda$-anomaly shifts towards higher temperatures with increasing applied field, this is particularly pronounced above the saturation field suggesting a Schottky anomaly.

\begin{figure}[htbp]
 \captionsetup{justification=raggedright, singlelinecheck=false}
 \includegraphics [width=\linewidth] {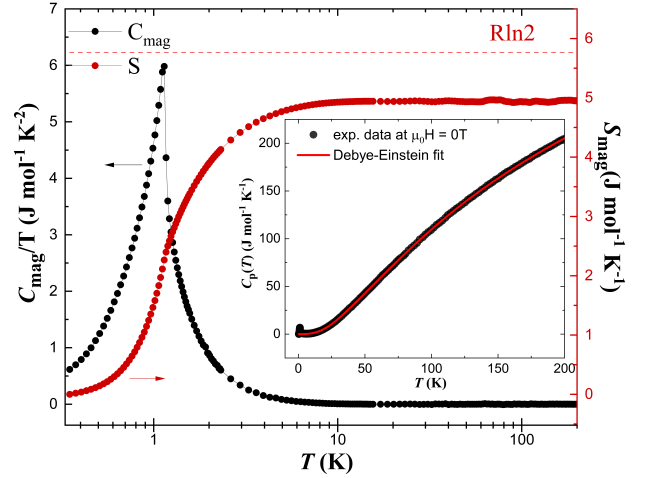}\\
 \caption{\justifying The inset shows the zero field heat capacity $C_{\rm p}$ vs $T$ in the temperature range 0.35-190~K. The solid red line represents the phonon contribution to the heat capacity $C_{\rm ph}$ obtained by fitting the raw data to Eq.~\ref{HC} in the 14-190~K temperature range. The main panel shows the magnetic heat capacity $C_{\rm mag}$ obtained by subtracting the $C_{\rm ph}$ from $C_{\rm p}$. $C_{\rm mag}/T$ is given by the black solid spheres (left y-axis). The solid red spheres shows the magnetic entropy $S_{\rm mag}$ (right y-axis) and the horizontal red dashed line represents the theoretical maximum entropy for a spin-$1/2$ system (Rln2 = 5.763~J/molK).}
 \label{entropy}
\end{figure}

In the case of a magnetic insulator, the zero field heat capacity consists of three components, the magnetic heat capacity due to spins $C_{\rm mag}(\textit{T})$, the lattice heat capacity arising from phonon vibrations $C_{\rm ph}(\textit{T})$, and the nuclear contribution, observable only at extremely low temperatures (below 0.1~K). In the absence of a nonmagnetic isostructural sample, the lattice contribution is determined by fitting $C_{\rm ph}(\textit{T})$ in the high-temperature range using the linear combination of one Debye and three Einstein modes as depicted in Eq.~\ref{HC} 
\begin{equation}\label{HC}
C_{\rm ph}(\textit{T}) = g_{\text{D}}C_{\text{D}} (\Theta_{D},T) + \sum_{i=1}^3g_{\text{E}_{i}}C_{\text{E}_{i}} (\Theta_{E},T) 
\end{equation}
Where $g_{D}$, $g_{\text{E}_{i}}$, are the respective weights of the Debye and three Einstein heat capacity contributions whose values are constrained such that their sum is equal to unity. The Debye ($C_{\text{D}}$) and Einstein ($C_{\text{E}}$) modes are given in Eq.~\ref{Deb} and Eq.~\ref{Ein} respectively.

\begin{equation}\label{Deb}
C_{\text{D}} = 9nR \left(\frac{T}{\Theta_{\text{D}}}\right)^3 \int_0^{\frac{\Theta_{\text{D}}}{T}} \frac{x^4 e^x}{(e^x - 1)^2} dx
\end{equation}

\begin{equation}\label{Ein}
C_{\text{E}_{i}} = 3nR \left(\frac{\Theta_{\text{E}_{i}}}{T}\right)^2 \frac{e^{\frac{\Theta_{\text{E}_{i}}}{T}}}{(e^{\frac{\Theta_{\text{E}_{i}}}{T}} - 1)^2}
\end{equation}
where n is the number of atoms in the formula unit, $R$ is the universal gas constant and $\Theta_{D}$, $\Theta_{E}$  are the Debye and Einstein temperatures respectively. The fitting was performed using one Debye and three Einstein terms in the temperature range $T_{l}$$<$$T$$<$195~K. The best fit (achieved for lower boundary of the temperature range, $T_{l}$$=$14~K) is shown in Fig.~\ref{entropy}, it yields $\Theta_{D}$ $\simeq$ 126~K, $\Theta_{\text{E}_{1}}$  $\simeq$ 181~K, $\Theta_{\text{E}_{2}}$  $\simeq$ 368~K, and
$\Theta_{\text{E}_{3}}$  $\simeq$ 525~K. 

The magnetic contribution to the heat capacity is obtained by extrapolating the lattice contribution down to low temperatures and subtracting it from the raw data. $C_{\rm mag}/T$ is shown in Fig.~\ref{entropy} and the corresponding change in magnetic entropy obtained by integrating over temperature is shown on the right y-axis. 
From Fig.~\ref{entropy}, the maximum entropy for the compound is 4.95~J/molK, which is lower than expected for a quantum spin-$1/2$ system (Rln2 = 5.763 ~J/molK). This could be due in part to the inadequacies of modelling the lattice contribution. A small fraction of entropy is also inevitably missed due to the limited lowest temperature of the experiment. However, this missing contribution can be estimated by fitting the low-temperature heat-capacity data below 1 K with a law of $C/T$$\sim$$T^{2}$ and extrapolating the fit to 0~K. Using this approach, we estimate that only about 0.07~J/molK of the entropy is not taken into account. Interestingly, a substantial part ($\approx 40-50$~\%) of the entropy is recovered above the transition, suggesting the presence of short-range magnetic correlations above $T_N$, as expected in a frustrated system.  

\subsection{Magnetic field versus temperature phase diagram}

\begin{figure}
 \includegraphics [scale=.32] {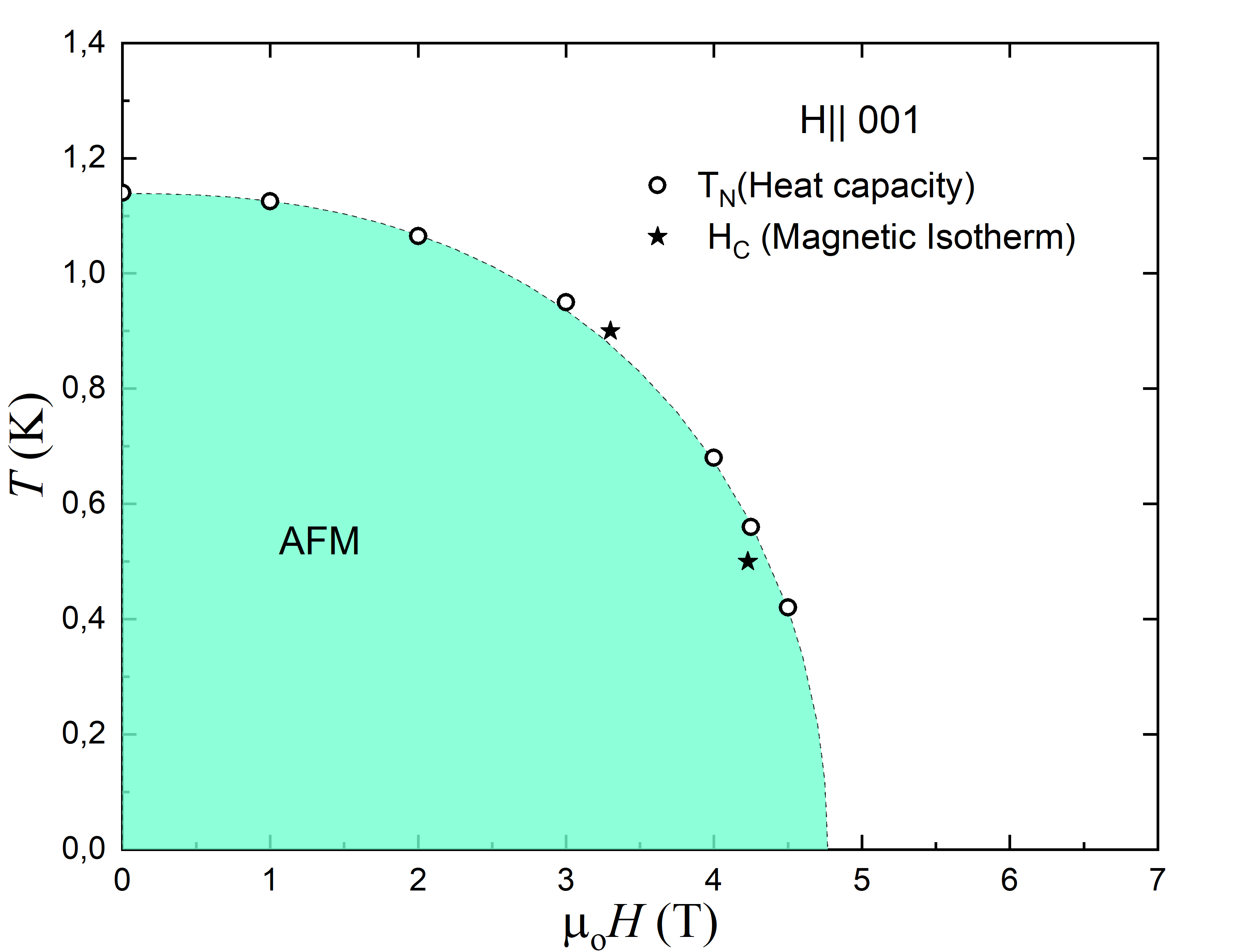}\\
 \captionsetup{justification=justified, singlelinecheck=false}
 \caption{\justifying The magnetic phase diagram as a function of magnetic field parallel to the c-axis versus temperature, constructed from heat capacity and magnetization experiments reveals a quantum phase transition from antiferromagnetic order to the polarized ferromagnetic state.}
 \label{Figphase}
\end{figure}

In order to visualize the magnetic properties of \LCGO, the phase diagram as a function of temperature versus magnetic field parallel to the {\bf c}-axis was constructed using heat capacity measurements in different fields and isothermal magnetization measurements at different temperatures. Figure~\ref{Figphase} suggests that the antiferromagnetic transition temperature is suppressed down to the absolute zero temperature at around 4.7~T above which the system becomes a polarized ferromagnet.

\subsection{Magnetic Structure determination}
 
\begin{figure}[htbp]
\includegraphics [scale=1.0] {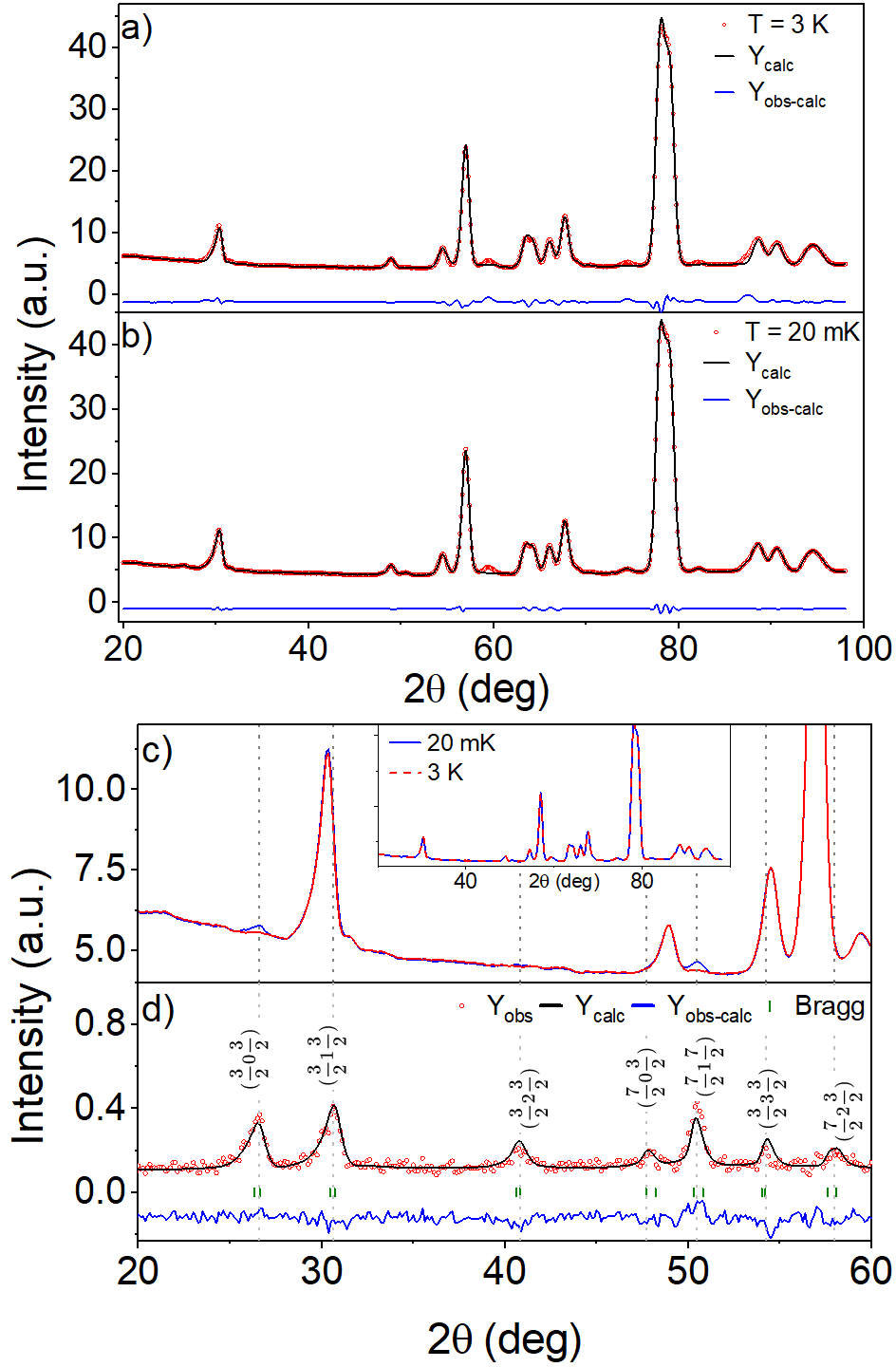}\\
\captionsetup{justification=justified, singlelinecheck=false}
\caption{\justifying a) The neutron powder diffraction at 3~K is plotted as open red circles, the rietveld refinement of the crystal structure is shown by the solid black line and the difference between the experiment and the calculation is given by the solid blue line. b) Neutron powder diffraction at 20mk (open red circles) is plotted with the rietveld refinement of the crystal and magnetic structure (solid black line) and the difference (solid blue line). c) Neutron powder diffraction data at 3~K~$>T_N$ (solid red line) and 20~mK~$\ll T_N$ (solid blue line) at low angles, the dotted dark Grey lines indicate the magnetic reflections, indexed in the format $G=H+\kappa$. The inset shows the data over the full angular range. d) The magnetic refinement (black line) of the difference between the 3~K and 20~mK datasets (open red circles) was achieved using the propagation vector $\kappa=(0.5,0,0.5)$ with a reliability factor of: $\chi^{2}$= 4.05.}
\label{figNPD}
\end{figure}

In order to determine the magnetic structure of \LCGO, Neutron Powder Diffraction was carried out at the high-intensity diffractometer WOMBAT at ANSTO. Neutron diffraction patterns were measured on the  powder sample synthesized by solid state reaction (rather than the crushed single crystal used for the X-ray measurements). Data was collected over the temperature range from 20~mK ($\ll T_N$) to 3~K ($> T_N$). The crystal structure is confirmed to be monoclinic with the $I1m1$ space group at these low temperatures, and thus there was no evidence of any structural phase transition down to 20~mK. Rietveld refinement of the neutron diffraction patterns at $T=3$~K and $T=20$~mK are shown in Fig.~\ref{figNPD}. A phase corresponding to the $I1m1$ space group explained the majority of the nuclear reflections at 3~K, but there are three additional temperature-independent reflections which were not present in the single crystal sample. These additional reflections could come from impurity phases in the solid state synthesized powder or from the instrument/dilution fridge. Two of the additional reflections at 2$\theta= 74.0$ deg and 2$\theta= 82.0$ deg correspond to the impurity phase La$_2$O$_3$, however the reflection at 2$\theta = 59.77$ deg did not fit any of the reported precursors or combinations of them, or any materials from the sample can/instrument setup. The peak at 2$\theta = 59.77$ deg was excluded from the refinement. The overall Rietveld fit yielded the proportion 97$\%$ of the \LCGO ~main phase and 3 vol.$\%$ of the La$_2$O$_3$ impurity in our powder sample. Figure \ref{figNPD}(a) shows the best fit of the crystal structure at 3K.

A comparison of the diffraction patterns obtained at varying temperatures indicates that the magnetic Bragg reflections emerge below $T=1.1$~K on cooling. The intensity of the magnetic reflections increases as the temperature decreases, confirming that \LCGO ~is indeed magnetically ordered below $T_N=1.1$~K in agreement with the measurements of susceptibility and heat capacity. The diffraction patterns at $T=3$~K and 20~mK are overplotted in Fig.\ref{figNPD}(c) and seven magnetic Bragg reflections are  observed 
as indicated by the dashed line. All magnetic reflections could be indexed with the commensurate propagation vector $\kappa=$~(0.5,0,0.5) using the software $k$-search of the \texttt{FullProf} \cite{Carvajal55} suite.

Due to the low intensity of the magnetic reflections, the refinement of the magnetic structure was performed taking into account the difference of the diffractograms measured at $T=3$~K and 20~mK.  Fig.\ref{figNPD}(d) shows this difference pattern along with the Rietveld refinement of the magnetic structure. The structural parameters for the magnetic structural refinements are fixed to the values obtained from the neutron powder refinement at 3~K. From the representation analysis the full decomposition of the magnetic representation in terms of irreducible representations (IRs) for the Wyckoff position 4b of $I1m1$ and propagation vector $\kappa=$~(0.5,0,0.5) are described by the equation.

\begin{equation}\label{IR}
\Gamma_{mag,Cu} = 3  \Gamma^{1}_{1}+3  \Gamma^{1}_{2},
\end{equation}

\begin{figure}[htbp]
\includegraphics [width=\linewidth] {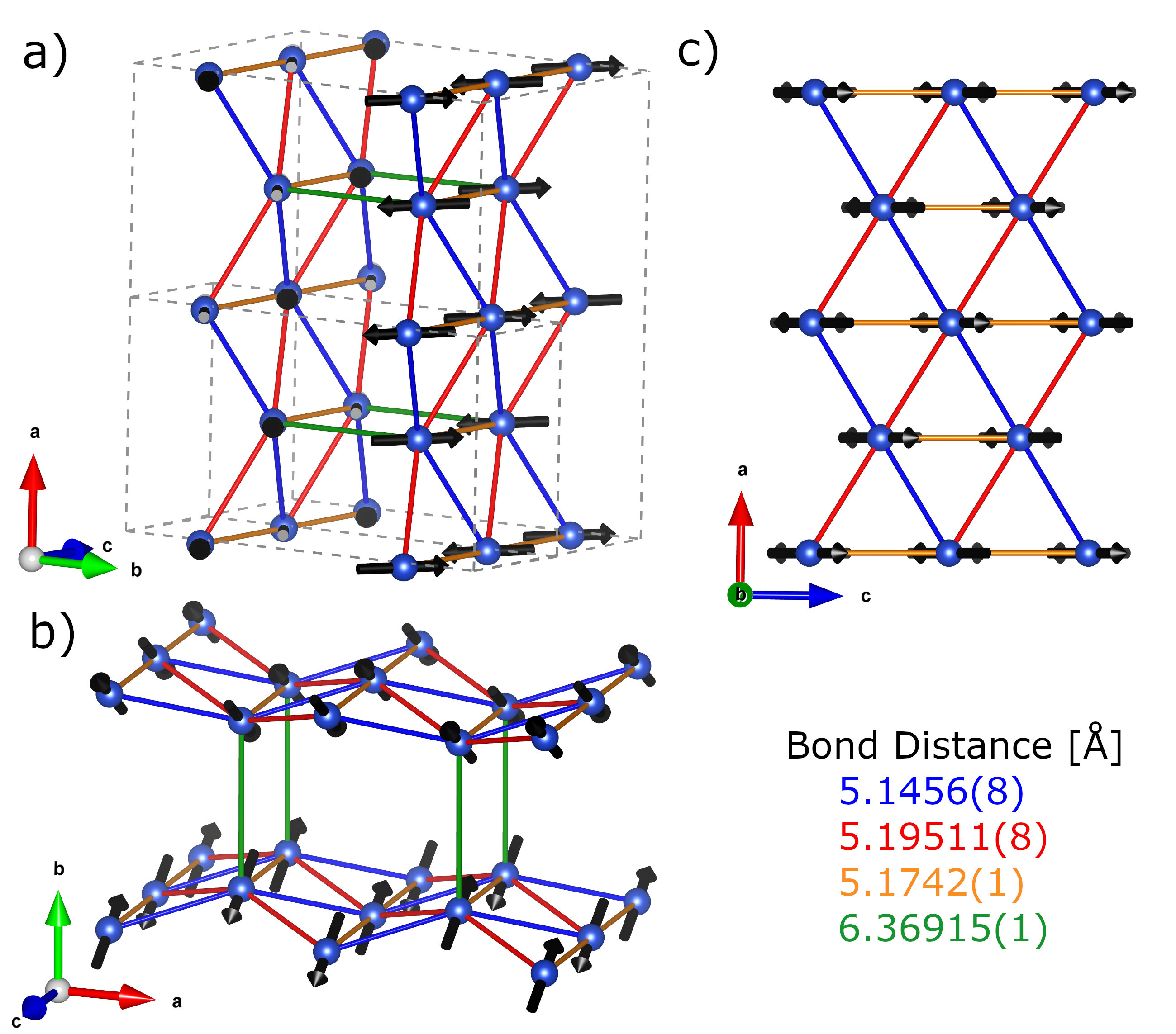}\\
\captionsetup{justification=justified, singlelinecheck=false}
\caption{\justifying Magnetic structure of \LCGO~showing the Cu$^{2+}$ ions only (blue spheres) while the black arrows indicate the ordering direction of their magnetic moments. a) shows one magnetic unit cell which is doubled along the {\bf a}- and {\bf c}-axes compared to the structural unit cell. The colored lines represent the different Cu$^{2+}-$~Cu$^{2+}$ bonds. In the {\bf ac}-plane, the bonds form buckled and distorted triangular layers with three different bond distances, the blue bond is the smallest with distance $d_{Cu-Cu}=5.1456(8)$~\AA, the orange bond has $d_{Cu-Cu}=5.1742(1)$~\AA~and the red bond has $d_{Cu-Cu}=5.19511(8)$~\AA. c) shows one triangular layer projected onto the {\bf ac}-plane over several unit cells. The triangular layers are stacked along the {\bf b}-axis with shortest distance $d_{Cu-Cu}=6.36915(0)$~\AA~given by the green bond. There are two layers per unit cell and b) shows these layers in a three-dimensional rendering of the magnetic structure. The spin directions are collinear within each triangular layer but are clearly non-collinear between layers. They lie in the {\bf bc} plane and are canted from the {\bf c}-axis by an angle of $\pm 33.1^{\circ}$.
}
\label{Fig7}
\end{figure}

Each of the terms corresponds to one of the irreducible representations ($t$$\Gamma^{d}_{o}$), where the coefficient $t$ gives the number of times it occurs, the subscript $o$ gives the order of the representation, and the superscript $d$ gives the dimensionality. Therefore, both $\Gamma_{1}$ and $\Gamma_{2}$ are one-dimensional. $\Gamma_{1}$ and $\Gamma_{2}$ were systematically tested in the magnetic structural refinement procedure using Fullprof \cite{Carvajal55}. The fitting was challenging because there are only a few low intensity magnetic reflections. Both irreps where tested and $\Gamma_1$ led to a satisfactory fit. The resulting magnetic structure has a reliability factor of: $\chi^{2}= 4.05$ and is depicted in Fig.~\ref{Fig7}.

The magnetic structure is described by the monoclinic $P_am$ space group (No. 6.21) where the unit cell doubled along the {\bf a}- and {\bf c}-axes and the Cu site splits into two, Cu1 (x,y,z) and Cu2 (x+1/4, y+1/2, z+3/4). Keeping in mind that the crystal structure is preserved to be monoclinic and there is in reality only one type of Cu ion in the structure, we have constrained the moments to be identical in both sites. Any deviation from equal moment magnitudes could not be statistically justified given the quality of the available data. The magnetic moments lie in the  {\bf bc}-plane with corresponding magnetic components of $m_b=0.50(3)$~$\mu_{B}$ and $m_c= 0.73(5)$~$\mu_{B}$ along the {\bf b}- and {\bf c}-axes respectively. The total ordered moment size of the Cu$^{2+}$ ions is therefore $M_{total}$= 0.89(6)$\mu_{B}/$Cu$^{2+}$ at 20~mK which matches the saturation values observed in magnetization and is close to the theoretically expected value of $M_{theory}= 1.0$~$\mu_{B}/$Cu$^{2+}$, for spin-1/2 ions. The observed spin direction with the spins perpendicular to the {\bf a}-axis is also consistent with the results of the magnetization measurements which observed spin-flop transitions for magnetic field parallel to the {\bf b}- and {\bf c}-axes but not for field parallel to the {\bf a}-axis.

Fig.~\ref{Fig7}(a) show the magnetic structure where the Cu$^{2+}$ ions are represented by the blue spheres and the magnetic moments are represented by the black arrows, the dotted lines depict the structural unit cell compared with the magnetic unit cell which is doubled on the {\bf a}- and {\bf b}- axis. Fig.~\ref{Fig7}(b) and (c) show the projection of the magnetic structure onto the {\bf ab}-, {\bf ac}- planes of the magnetic unit cell respectively,  The colored lines indicate the different Cu$^{2+}$-Cu$^{2+}$ distances corresponding to inequivalent exchange interactions. The triangular layers form in the {\bf ac}-plane, and Fig.~\ref{Fig7}(c) shows the projection of one triangular layer onto this plane over several unit cells. The triangular layers are distorted and there are three non-equivalent interactions represented by the blue, orange, and red lines corresponding to bond distances of $5.1456(8)$~\AA, $5.1742(1)$~\AA\ and $5.19511(8)$~\AA. The triangular layers are stacked along the {\bf b}-axis with two layers per unit cell and with a smallest interlayer distance of $6.36915(0)$~\AA~represented by the green bond. Fig.~\ref{Fig7}(b) shows these layers in a three-dimensional rendering of the structure. It is clear that the layers are buckled rather than flat.

Within each triangular layer the spins are collinear. They lie within the {\bf bc}-plane and are canted by an angle of $33.1^{\circ}\pm1^{\circ}$ from the {\bf c}-axis. The arrangement is antiferromagnetic with each spin canceled by a spin of opposite direction. Within each layer the spins form antiferromagnetic chains coupled by the orange and blue bonds (best seen Fig.~\ref{Fig7}(c)) suggesting that these interactions are antiferromagnetic.  The spins coupled by the red bonds are parallel, suggesting that this interaction which is probably antiferromagnetic is the weakest and does not influence the structure. Neighboring triangular layers stacked along the {\bf c}-axis have opposite canting direction possibly due to competing interlayer interactions, thus overall the magnetic structure is co-planar but non-collinear.

\section{\textbf{Summary}}
We successfully synthesized large high-quality single crystals of the spin-$1/2$ distorted triangular lattice compound CuLa$_2$Ge$_2$O$_8$ using the TSFZ technique. Previously, sub-millimeter-sized crystals were grown by the flux method. The phase purity and excellent crystallinity of our as-grown crystals were rigorously verified. The critical factors for the successful growth of this compound included careful selection of an appropriate solvent ratio, precise control of the growth atmosphere, and the use of a slow growth rate. 

Magnetic and thermodynamic measurements show that CuLa$_2$Ge$_2$O$_8$ orders at $T_{N} = 1.14(1)$~K. The Curie-Weiss fitting yields a Curie-Weiss temperature of -3.74~K, revealing antiferromagnetic interactions and implying possible weak frustration. The magnetization data showed a behavior characteristic of a low-dimensional antiferromagnet at low temperatures. An applied magnetic field suppressed the antiferromagnetic order and achieved saturation at $\mu_0 H = 4.25$~T. It should be noted that we could not see the field-induced transitions at $H_{C1} = 0.08$~T and $H_{C2} = 0.58$~T as previously reported \cite{cho2017properties}, however, we observed a spin-flop transition at $\mu_0H=0.42$~T for certain field directions, implying that the spins point in the {\bf bc} plane. We also observed a field-dependent Schottky anomaly that was previously unreported. 

The magnetic structure of this compound was investigated using neutron powder diffraction and found to be a co-planar antiferromagnetic state with the spins pointing in the {\bf bc}-plane but different from the 120$^{\circ}$ order of an undistorted triangular antiferromagnet.

The total ordered moment is M$_{total}$= 0.89(6)$\mu_{B}/$Cu$^{2+}$ at 20~mK. Having optimized the growth parameters and successfully synthesized high-quality single crystals, our next objective is to conduct more comprehensive magnetic characterizations, including inelastic neutron scattering experiments and in-depth theoretical analysis to gain deeper insights into the quantum magnetism of the system.

\section{\textbf{Acknowledgement}}
This research is supported by the Deutsche Forschungsgemeinschaft (DFG) through the project B06 of the SFB 1143 (ID 247310070). We acknowledge the support of the Australian Nuclear Science and Technology Organization in providing neutron research facilities used in this work. The authors also acknowledge the Core Lab Quantum Materials, HZB, Germany, where the powder and single-crystal samples were synthesized and the bulk thermodynamic and magnetic properties were measured.

\bibliography{document}

\end{document}